\documentclass[a4paper,11pt]{article}
\usepackage{pos}
\usepackage{amsmath,amsfonts,ulem,graphicx,footmisc,soul,enumitem,bbm,nicefrac,float,booktabs,array,mwe,dsfont,bm,citesort}
\usepackage{mathtools, slashed}
\usepackage[utf8]{inputenc}

\title{Towards the finite-volume spectrum of the Roper resonance}
\ShortTitle{Towards the finite-volume spectrum of the Roper resonance}

\author*[a]{Daniel Severt}

\affiliation[a]{Helmholtz-Institut f\"ur Strahlen- und Kernphysik and Bethe Center for Theoretical Physics, \\
Universit\"at Bonn, D-53115 Bonn, Germany}

\emailAdd{severt@hiskp.uni-bonn.de}

\abstract{The finite-volume energy levels corresponding to the Roper resonance based on a two-flavor chiral effective Lagrangian at leading one-loop order are investigated. It is shown that the Roper mass can be extracted from these levels for not too large lattice volumes. Further, to include three-body $N \pi \pi$ dynamics, a non-relativistic effective field theory for the Roper resonance within a covariant particle-dimer picture is introduced. This particle-dimer approach is a suitable framework to investigate three-particle scattering relevant for the Roper channel. The appearing dimer fields are analyzed, the energy levels of the Roper resonance in a finite volume are calculated and compared to the results from the chiral effective Lagrangian.}

\FullConference{%
The 39th International Symposium on Lattice Field Theory,\\
8th-13th August, 2022,\\
Rheinische Friedrich-Wilhelms-Universität Bonn, Bonn, Germany
}

\begin{document}
\maketitle
	
\section{Introduction}

Over the last years a lot of effort has been made to deepen our understanding of the hadron spectrum from QCD (for a current state of the art review see~\cite{Mai:2022eur}). Especially, the excited baryon spectrum remains puzzling up to this day. One of the most prominent and challenging systems is the Roper resonance. Discovered in $1964$ via partial wave analysis of $N \pi$-scattering data~\cite{Roper:1964zza}, the Roper, or $N(1440)$, posseses some interesting features. The resonance has identical quantum numbers as the nucleon, $I (J^P) = \nicefrac{1}{2} (\nicefrac{1}{2}^{+})$, but a larger mass (pole mass: $m_R = 1.365\,$GeV~\cite{PDG}). The most remarkable characteristic of the Roper lies in its decays: It can decay into a nucleon and a pion, as well as into a nucleon and two pions (via $\Delta \pi$ and $N\sigma$ intermediate states), where the branching ratios of these two decay modes are of the same magnitude. This makes two- and three-particle final states equally likely and complicates the Roper system significantly.

There are several options to explore the spectrum of the Roper resonance. One method is Lattice QCD (LQCD), a non-perturbative approach to QCD, where numerical calculations of correlation functions are performed on a discretized Euclidean space-time in a finite volume. There are already some preliminary studies of the Roper system in LQCD (see~\cite{Lang:2016hnn,Liu:2016uzk}), which indicate that both two-particle $N \pi$ and three-particle $N \pi \pi$ dynamics might be important to generate the Roper resonance. Another method is an effective field theory (EFT) approach for the Roper, which has already been established within the framework of baryon chiral perturbation theory (BChPT), see Refs.~\cite{Gegelia:2016xcw,Beane:2002ud,Borasoy:2006fk,Djukanovic:2009gt,Long:2011rt}. Here, the Roper resonance is included in BChPT as an explicit degree of freedom and one can calculate various observables, like decay widths, within this framework. However, difficulties arise in general when comparing results from EFT with results from LQCD. For example, the fact that LQCD calculations are always performed in a box of finite size, causes a shift in the energy levels of the system. In order to improve the investigation of the Roper resonance and to simplify the comparison between LQCD and infinite volume EFT, one can use an EFT approach in a finite volume (see e.g.~\cite{Meissner:2022odx}). When a narrow resonance is present, the energy levels show a particular behaviour near the resonance's energy: The levels shift when the box size $L$ is changed, but they do not cross each other. This behaviour is known as ``avoided level crossing''~\cite{Wiese:1988qy} and it has been discovered in other resonance systems, such as the delta resonance~\cite{Bernard:2007cm}.

For the Roper resonance, such a study has been done in Ref.~\cite{Severt:2020jzc}, where the two-particle final states $N \pi$ and $\Delta \pi$ have been considered, but not yet three-particle final states. The two-body sector in a finite volume is already well established from L\"uscher's method~\cite{Luscher:1986pf,Luscher:1990ux}, whereas the three-body sector remains challenging. A lot of work has been done and different methods have been introduced to tackle the three-particle dynamics. These methods usually require the formulation of a so-called three-body quantization condition (for some examples, see Refs.~\cite{Mai:2021lwb,Mai:2017bge,Briceno:2017tce,Doring:2018xxx,Sharpe:2017jej,Guo:2016fgl,Hansen:2016ync,Hansen:2015zga,Meissner:2014dea,Muller:2020wjo}). A very promising approach to describe three-particle scattering in a finite box is the particle-dimer framework (see~\cite{Bedaque:1998kg,Bedaque:1998km,Braaten:2004rn,Hammer:2017kms,Hammer:2017uqm}), which reformulates the three-body problem as a two-body problem and, therefore, significantly simplifies the three-particle dynamics. This particle-dimer picture might also be a suitable approach to investigate the Roper resonance.

In this talk, the main results of Ref.~\cite{Severt:2020jzc} are summarized and a perspective is given how these findings can be improved. After that, a non-relativistic particle-dimer approach for the Roper resonance is introduced to include explicitely three-particle $N \pi \pi$ dynamics. 

\clearpage

\section{The Roper resonance in BChPT}

\begin{figure}[t]
	\centering{
		\includegraphics*[width=0.8\linewidth, trim=0 .0cm 0 0]{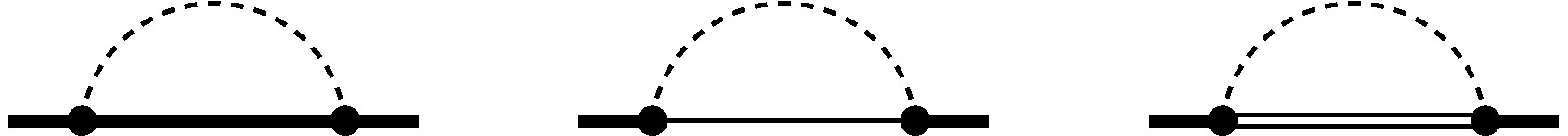} 
	}
	\caption{
		One-loop diagrams contributing to the Roper mass at third chiral order. 
		Thick solid, dashed, solid, and solid double lines refer to the Roper resonance, pions, nucleons, and delta
		baryon states, respectively. The vertices denoted by a filled dot refer to insertions from the first order
		chiral Lagrangian.
		\label{fig:diags1}
	}
\end{figure}

First of all, let us consider how the Roper resonance can be treated in BChPT and how its energy levels can be obtained in a finite volume. The effective Lagrangian for the Roper system is given by  
\begin{align}
\mathcal{L}_{\text{eff.}} = \mathcal{L}_{\pi \pi} + \mathcal{L}_{\pi N} + \mathcal{L}_{\pi R} + \mathcal{L}_{\pi \Delta}
+ \mathcal{L}_{\pi N \Delta} + \mathcal{L}_{\pi N R} + \mathcal{L}_{\pi \Delta R} \; , 
\end{align}
where the dynamical degrees of freedom are pions ($\pi$), nucleons ($N$), the delta ($\Delta$) and the Roper resonance ($R$). The Lagrangian is restricted to flavor $SU(2)$ and the isospin limit ($m_u = m_d = \hat{m}$) is used throughout~\cite{Severt:2020jzc}. Additionally, the calculations are limited to third chiral order, $\mathcal{O}(p^3)$, where $p$ denotes a momentum or mass much smaller than $\Lambda_{\chi}$, the chiral-symmetry breaking scale ($\Lambda_{\chi} \approx 1\,$GeV). The chiral power counting is discussed detailetly in Refs.~\cite{Gegelia:2016xcw,Severt:2020jzc}. The different particles can interact among themselves and induce self-energy corrections. This is also the case for the Roper resonance and the leading one-loop diagrams are given in Fig.~\ref{fig:diags1}. There are also contact interactions contributing to the self-energy, which are not shown here. 
The three loop diagrams differ by the internal baryon state, which can be either a Roper, a nucleon, or a delta baryon. All these self-energy contributions generate a correction to the Roper resonance propagator and ``dress'' it, i.e. 
\begin{align}
i S_{R} \left( p \right) = \frac{i}{\slashed{p} - m_{R 0} - \Sigma_{R} \left( \slashed{p} \right) } \; ,  
\end{align}
where $m_{R 0}$ denotes the bare mass of the Roper and $\Sigma_{R}$ is the self-energy. From this equation, the mass of the Roper can be calculated by evaluating the self-energy and finding the pole of the propagator. In the infinite volume, the self-energy can be calculated by standard methods in BChPT (see e.g.~\cite{Severt:2019sbz}). In the finite volume, on the other hand, the loop integral over the spatial momenta is replaced by an infinite sum of quantized momenta while the integration over the (Euclidean) time component remains unchanged 
\begin{align}
\int \frac{d^4 k_E}{(2 \pi)^4} \left(\ldots \right) \mapsto \int_{- \infty}^{+ \infty} \frac{d k_4}{2 \pi} \frac{1}{L^3} \sum_{\vec{k}}
\left(\ldots \right) \; , \quad \text{with} \quad \vec{k} =  \frac{2 \pi}{L} \vec{n} \; , \quad \vec{n} \in \mathbb{Z}^3 \; . \label{FV-formalism}
\end{align}
This change obviously influences the self-energy and propagator of the Roper resonance, whose poles are now given by 
\begin{align}
\slashed{p} - m_{R0}^{}  - \Sigma_{R}^L \left( \slashed{p} \right) = 0 \; , \label{FVpoles-ChPT}
\end{align}
where $\Sigma_{R}^L \left( \slashed{p} \right)$ denotes the self-energy of the Roper in the finite box. Choosing the rest frame of the Roper, i.e. $p = ( E, \vec{0} )$, and using the on-shell condition $\slashed{p} = E$, one obtains a relation for the finite-volume energy levels from Eq.~\eqref{FVpoles-ChPT}  
\begin{align}
E - m_R - \tilde{\Sigma}_{R}^L \left( E \right) = 0 \; . 
\end{align} 
Here, $m_{R} = m_{R 0} + \text{Re} \left \lbrace \Sigma_{R}^{} \left( E \right) \right \rbrace$ denotes the physical Roper mass and $\tilde{\Sigma}_{R}^L \left( E \right)$ is the so-called finite-volume correction of the Roper self-energy, i.e. the difference between the finite volume and infinite volume self-energy contribution, $\tilde{\Sigma}_{R}^L \left( E \right) := \Sigma_{R}^L \left( E \right) - \text{Re} \left \lbrace \Sigma_{R}^{} \left( E \right) \right \rbrace$. Now, to obtain the energy levels one has to evaluate the different loop diagrams from Fig.~\ref{fig:diags1} in the finite and infinite volume and subtract them from each other. After neglecting all exponentially suppressed contributions the equation for the energy levels reads~\cite{Severt:2020jzc}
\begin{align}
m_{R}^{} - E &= \frac{3 g_{\pi N R}^2}{128 \pi^2 F_{\pi}^2 E} \left( E+ m_N^{} \right)^2 \left[ \left( E - m_N^{} \right)^2 -
M_{\pi}^2 \right] \tilde{B}_0^L \left( E^2, m_N^2, M_{\pi}^2 \right) \notag \\ 
&\phantom{=} \; + \frac{h_{R}^2}{96 \pi^2 F_{\pi}^2 m_{\Delta}^2 E} \left[ \left( m_{\Delta}^{} + E \right)^2 - M_{\pi}^2 \right]
\lambda \left( E^2, m_{\Delta}^{2}, M_{\pi}^{2} \right) \tilde{B}_0^L \left( E^2, m_{\Delta}^2, M_{\pi}^2 \right) \; , \label{EnLevels-ChPT}
\end{align}
with the nucleon mass $m_N$, the delta mass $m_{\Delta}$, the pion mass $M_{\pi}$, the pion decay constant $F_{\pi}$ and the Roper-pion-nucleon and Roper-pion-delta coupling constants $g_{\pi N R}$ and $h_R$, respectively. The function $\tilde{B}_0^L \left( E^2, m_{X}^2, M_{\pi}^2 \right)$ contains the finite volume sums, similar to the L{\"u}scher function~\cite{Luscher:1990ux}. One can observe that the loop functions with internal nucleons and internal delta baryons contribute to the energy levels, whereas the loop diagram with internal Roper generates only exponentially suppressed contributions, which are neglected. 

\begin{figure}[t]
	\includegraphics[width=0.49\linewidth]{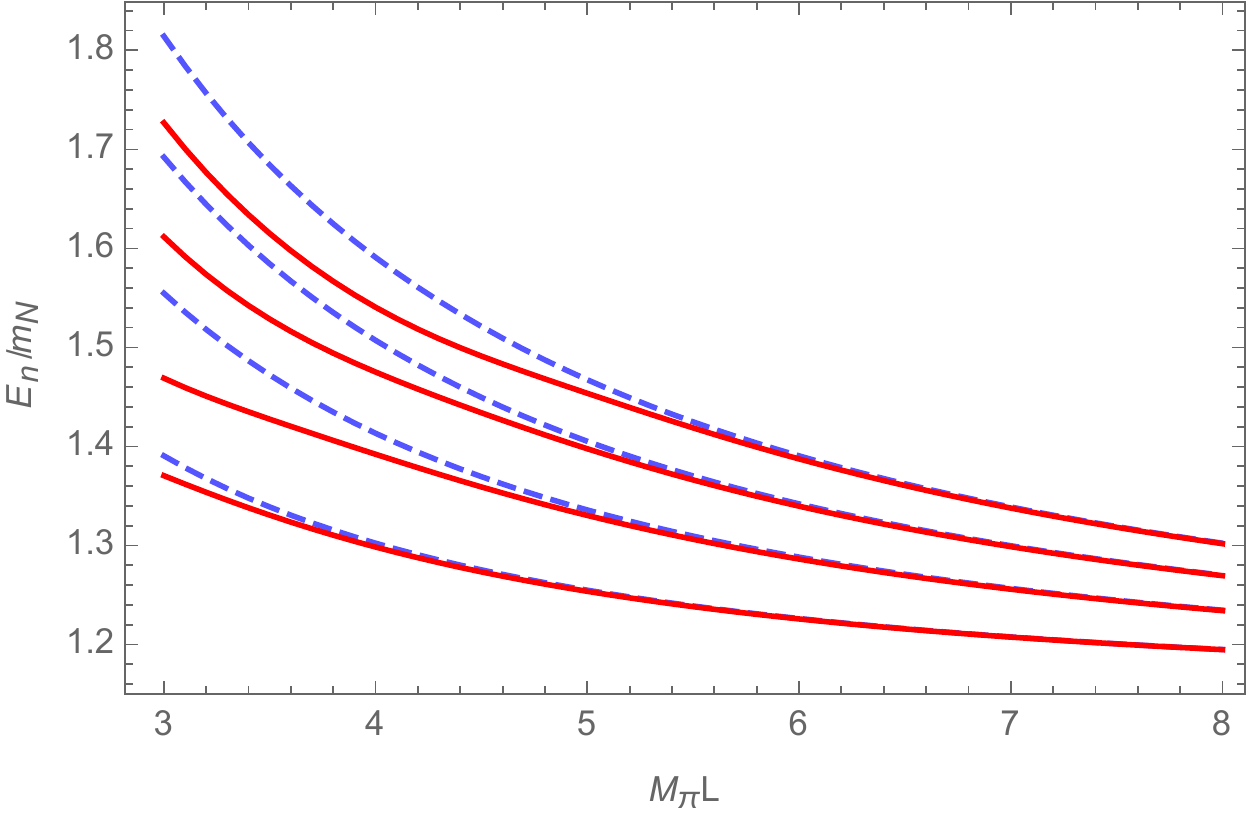}
	~
	\includegraphics[width=0.49\linewidth]{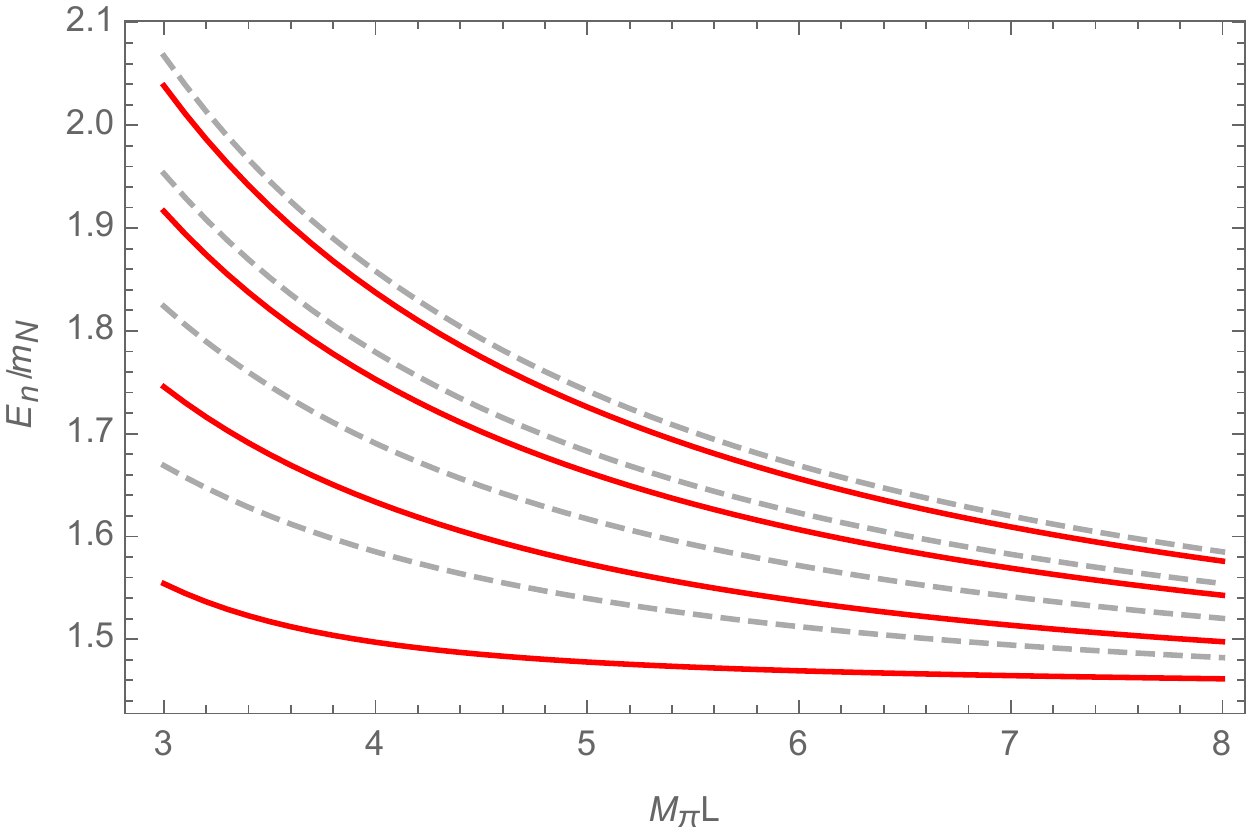}
	\caption{
		\textbf{Left}: Energy levels of the Roper resonance for different box sizes $L$ considering only pion and nucleon as intermediate states. Red solid lines display the numerical results for the interacting levels and blue dashed lines the free (non-interacting) energy levels of the pion and nucleon. 
		\textbf{Right}: Energy levels of the Roper resonance for different box sizes $L$ considering only pion and delta baryon as intermediate states. Red solid lines display the numerical results for the interacting levels and gray dashed lines the free (non-interacting) energy levels of the pion and delta.}
	\label{fig:En-levels-Npi-Ndelta}
\end{figure}

Now, Eq.~\eqref{EnLevels-ChPT} can be evaluated numerically to obtain the energy levels for different box sizes $L$. To simplify the matter, it is useful to look at the different contributions isolated: First, only the loop with internal nucleon is considered ($h_R=0$), then, only the loop with internal delta ($g_{\pi N R}=0$). The obtained energy levels for both cases are displayed in Fig.~\ref{fig:En-levels-Npi-Ndelta}. In the $N \pi$ case (Fig.~\ref{fig:En-levels-Npi-Ndelta}, left) one can observe a clear sign of avoided level crossing around the Roper mass $1.365\,$GeV$/m_N \approx 1.45$, which is from here on called the ``critical value''. For larger box sizes ($M_{\pi} L \approx 6$), the energy levels asymptotically approach the non-interacting (free) pion-nucleon energy levels, as one would expect. In the $\Delta \pi$ case (Fig.~\ref{fig:En-levels-Npi-Ndelta}, right) there are no signs of avoided level crossing. One reason for this might be that the energy levels lie well above the critical value, in fact only the two lowest levels come close to it. Additionally, the value of $h_R$ is quite large~\cite{Severt:2020jzc}, which tends to ``wash out'' the typical avoided level crossing signature (see e.g.~\cite{Bernard:2007cm}).  

\begin{figure}[t]
	\includegraphics[width=0.49\linewidth]{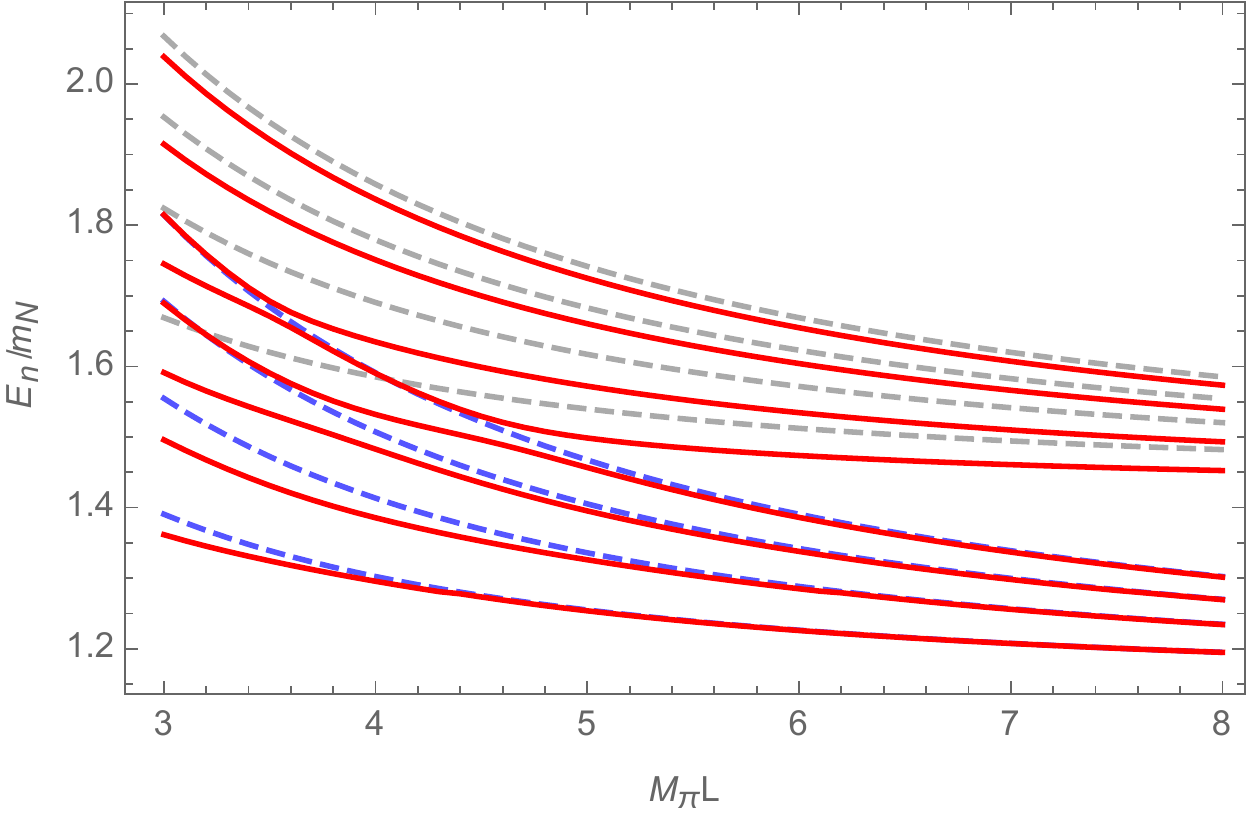}
	~
	\includegraphics[width=0.49\linewidth]{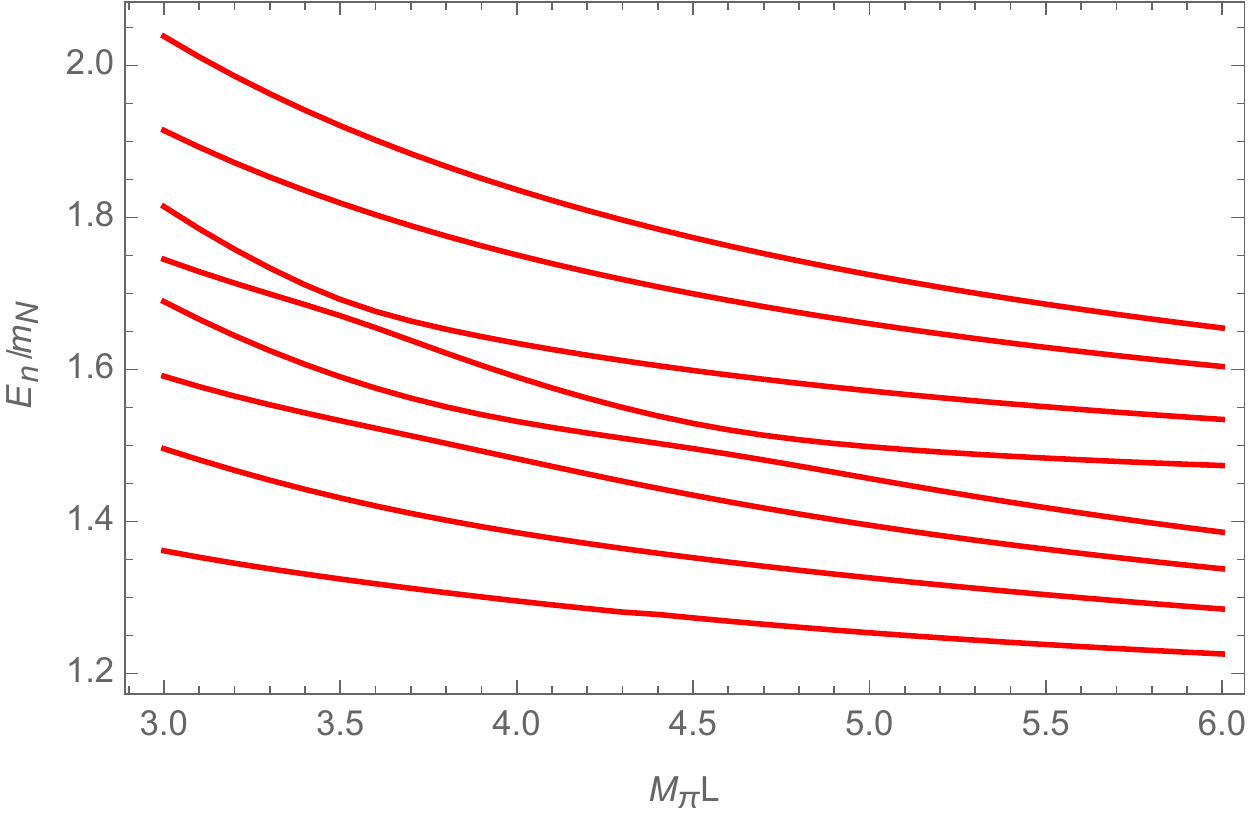}
	\caption{
		\textbf{Left}: Energy levels of the coupled system $N \pi$ and $\Delta \pi$ for different box sizes $L$. Red solid lines display the numerical results, blue dashed lines display the free energy levels of the pion and nucleon, and gray dashed lines display the free energy levels of the pion and delta baryon.  
		\textbf{Right}: Energy levels of the coupled system for different box sizes $L$ without displaying the non-interacting levels.}
	\label{fig:En-levels-coupledchannel}
\end{figure}

Next, both possible interactions are turned on and the energy levels of the coupled-channel system are calculated. The obtained spectrum is shown in Fig.~\ref{fig:En-levels-coupledchannel}. One observes again signs of avoided level crossing around the critical value $1.45$, where also many energy levels lie very close to each other. The signs fade out for energies further away from the critical value and for box sizes larger than $M_{\pi} L = 6$. For a better display of the energy level curves, see the right plot in Fig.~\ref{fig:En-levels-coupledchannel}, which does not depict the non-interacting levels.

This calculation demonstrates how the Roper resonance can be treated with BChPT in a finite volume. The two possible final states $N \pi$ and $\Delta \pi$, where the delta baryon is considered to be stable, are the most important decay channels in this framework. However, the delta resonance is of course not a stable particle and decays most likely into a nucleon and a pion, which results in the typical $N \pi \pi$ final state for the Roper. To explicitely include three particle dynamics into the system, a different approach needs to be considered.

\section{Particle-dimer formalism for the Roper resonance}

The particle-dimer picture was introduced to provide a convenient treatment of three-particle dynamics (see e.g. Refs.~\cite{Hammer:2017kms,Hammer:2017uqm} for a detailed introduction). The main idea is to introduce an auxiliary field, the so-called ``dimer field'', into the theory, which incorporates two-particle scattering information. Then, instead of calculating three-particle scattering, one calculates the scattering of one particle with the dimer field. The obtained particle-dimer scattering amplitude is equivallent to the original three-particle scattering amplitude~\cite{Bedaque:1998kg,Braaten:2004rn}. The goal is to use this particle-dimer formalism also for the Roper resonance.

To describe a three-particle $N \pi \pi$ system, consider the following non-relativistic EFT described by the Lagrangian $\mathcal{L}_{\pi \pi N}$ containing non-relativistic nucleon fields ($\psi$) and non-relativistic pion fields ($\phi$) 
\begin{align}
\begin{split}
\mathcal{L}_{\pi \pi N} = \mathcal{L}_{dyn.} &+ c_1 \phi^{\dagger}  \phi^{\dagger} \phi \phi + c_2 \psi^{\dagger} \phi^{\dagger} \phi \psi + c_3 \psi^{\dagger} \phi^{\dagger} ( \phi^{} + \phi^{\dagger} ) \phi \psi + c_4 \psi^{\dagger} \phi^{\dagger} \phi^{\dagger} \phi^{} \phi \psi + \ldots  \; . 
\end{split} \label{piN-Lagrangian}
\end{align} 
The interaction between these non-relativistic particles is parametrized by the low-energy constants (LECs) $c_{1,2,3,4}$ and the ellipses denote terms with higher numbers of (pion) field insertions and terms with derivatives, which are not taken into account for now. The first term in the Lagrangian, $\mathcal{L}_{dyn.}$, describes the covariant dynamics of the fields~\cite{Colangelo:2006va,Bernard:2008ax}
\begin{align}
\begin{split}
\mathcal{L}_{dyn.} &= \mathcal{L}_{\phi} + \mathcal{L}_{\psi} = \phi^{\dagger} 2 W_{\pi} \left( i \partial_t - W_{\pi} \right) \phi + \psi^{\dagger} 2 W_{N} \left( i \partial_t - W_{N} \right) \psi \; , 
\end{split} \label{piN-dyn-Lagrangian}
\end{align}
with the differential operators $W_{\pi} = \big[ M_{\pi}^2 - \vec{\nabla}_{\phantom{\pi}}^2  \big]^{1/2}$ and $W_{N} = \big[ m_{N}^2 - \vec{\nabla}_{\phantom{N}}^2  \big]^{1/2}$, where $M_{\pi}$ is the pion mass and $m_{N}$ the nucleon mass. The square root structure of these operators ensures that resulting amplitudes will be relativistically invariant and leads to the known relativistic energy-momentum relation in momentum space, which can be seen in the propagators for the nucleon and the pion 
\begin{align}
S_{N} \left( p_0, \vec{p} \right) &= \frac{1}{2 \omega_{N} (\vec{p}) \left[  \omega_{N} (\vec{p}) - p_0 - i \epsilon \right]} \; , \quad S_{\pi} \left( p_0, \vec{p} \right) = \frac{1}{2 \omega_{\pi} (\vec{p}) \left[ \omega_{\pi} (\vec{p}) - p_0 - i \epsilon \right]} \; , 
\end{align} 
where $\omega_{N} (\vec{p}) = \big[ | \vec{p} |^{2} + m_{N}^{2} \big]^{1/2}$ and $\omega_{\pi} (\vec{p}) = \big[ | \vec{p} |^{2} + M_{\pi}^{2} \big]^{1/2}$. 

Now, that the dynamics and interactions of the nucleon- and pion-fields are known, it is time to introduce the dimer fields. Since there are two different fields (nucleon and pion), a dimer field that accounts for pion-pion scattering, as well as a dimer field that accounts for nucleon-pion scattering, is needed. In total, three dimer fields have to be included: The first dimer field is $\Delta$, which has the quantum numbers of the spin-$3/2$ delta resonance ($J^P = \nicefrac{3}{2}^{+}$) and incorporates intermediate nucleon-pion interactions. The second dimer field is $\sigma$ with the quantum numbers of the scalar isoscalar resonance, the $f_0 (500)$ ($J^P = 0^{+}$), also known as $\sigma$ meson, which accounts for intermediate pion-pion interactions. Note that the quantum numbers of the delta resonance and pion system and the $f_0 (500)$ and nucleon system overlap with the quantum numbers of the Roper resonance. Finally, the third dimer field $R$ is for the Roper resonance itself. It has the same quantum numbers as the nucleon ($J^P = \nicefrac{1}{2}^{+}$) but a larger mass $m_R$. Considering all these dimer fields, one obtains the Lagrangian 
\begin{align}
\begin{split}
\mathcal{L}_{Dimer} = \mathcal{L}_{dyn.} + \mathcal{L}_{T} \; , \label{Dimer-Lagrangian}
\end{split}
\end{align} 
where $\mathcal{L}_{dyn.}$ is taken from Eq.~\eqref{piN-dyn-Lagrangian} and $\mathcal{L}_{T}$ contains the dimer fields and their interactions 
\begin{align}
\begin{split}
\mathcal{L}_{T} &= R^{\dagger} 2 W_{R} \left( i \partial_t - W_{R} \right) R + f_1 R^{\dagger} \phi^{\dagger} \phi^{} R - f_2 [ R^{\dagger} \phi^{} \psi^{} + R^{} \phi^{\dagger} \psi^{\dagger} ] \\ 
& \phantom{=} - f_3 [ R^{\dagger} \phi \Delta + \Delta^{\dagger} \phi^{\dagger} R ] - f_4 [ R^{\dagger} \sigma \psi + \psi^{\dagger} \sigma^{\dagger} R ] \\ 
& \phantom{=} + \alpha_{\Delta}^{} m_{\Delta}^2 \Delta^{\dagger} \Delta + g_1 \Delta^{\dagger} \phi^{\dagger} \phi^{} \Delta - g_2 [ \Delta^{\dagger} \phi^{} \psi^{} + \Delta^{} \phi^{\dagger} \psi^{\dagger} ] \\
& \phantom{=} + \alpha_{\sigma}^{} M_{\sigma}^2 \sigma^{\dagger} \sigma + h_1 \psi^{\dagger} \sigma^{\dagger} \sigma \psi - h_2 [ \sigma^{\dagger} \phi \phi + \sigma^{} \phi^{\dagger} \phi^{\dagger} ] \\ 
& \phantom{=} - G_{R \sigma} [ R^{\dagger} \phi^{\dagger} \sigma \psi + \psi^{\dagger} \sigma^{\dagger} \phi R ] - G_{R \Delta} [ R^{\dagger} \phi^{\dagger} \phi \Delta + \Delta^{\dagger} \phi^{\dagger} \phi R ] \\ 
& \phantom{=} - G_{\Delta \sigma} [ \Delta^{\dagger} \phi^{\dagger} \sigma \psi + \psi^{\dagger} \sigma^{\dagger} \phi \Delta ] \; . \label{Dimer-T-Lagrangian}
\end{split}
\end{align} 
The interactions among the different fields are described by the LECs $f_{1,2,3,4}$, $g_{1,2}$, $h_{1,2}$, and $G_{R \sigma, R \Delta, \Delta \sigma}$. For simplicity, interaction terms with more field insertions and with derivatives are not included in this pioneering work. Also, spin and isospin projections are not considered for now. There are some things to note from Eq.~\eqref{Dimer-T-Lagrangian}: First, the Roper-dimer field $R$ has the same dynamical Lagrangian as the pion and the nucleon, whereas the $\Delta$- and $\sigma$-dimer are static, i.e. the Lagrangian does not contain time or spatial derivatives of these fields. The fact that $R$ is dynamic should ensure a more accurate treatment of the Roper's properties. Second, the dimer fields can interact among themselves. For example, the Roper-dimer can decay into a nucleon and a $\sigma$-dimer, or a pion and a $\Delta$-dimer. Integrating out the $\sigma$- and and a $\Delta$-dimer, one obtains a Lagrangian containing only the Roper resonance, nucleons and pions and vertices appear that allow the decay of Roper into $N \pi$ and $N \pi \pi$ final states. If one integrates out also the Roper-dimer, one arrives again at Eq.~\eqref{piN-Lagrangian}.

\begin{figure}[t]
	\centering{
		\includegraphics*[width=0.9\linewidth, trim=0 .0cm 0 0]{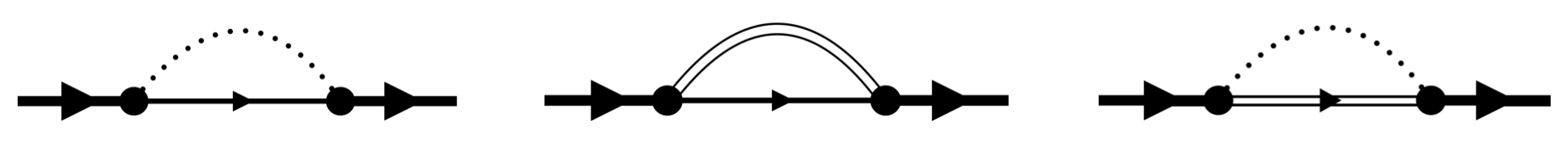} 
	}
	\caption{
		Feynman diagrams contributing to the Roper resonance mass at one-loop order. The thick solid line with arrow, the solid line with arrow and the double solid line with arrow refer to the Roper resonance, nucleons, and $\Delta$-dimer fields, respectively. The dotted line represents pions and the double solid line $\sigma$-dimer fields. 
		\label{fig:diags2}
	}
\end{figure}

Analogously to the BChPT investigation (see last paragraph), one can now take a look at the self-energy corrections of the Roper-dimer field. From the interactions in Eq.~\eqref{Dimer-T-Lagrangian}, three Feynman diagrams appear at one-loop order, which are depicted in Fig.~\ref{fig:diags2}. The diagrams differ again by their intermediate states: A nucleon and a pion, a nucleon and a $\sigma$-dimer, or a pion and a $\Delta$-dimer can appear inside the loop. The dressed Roper resonance propagator in the non-relativistic EFT is given by 
\begin{align}
S_{R}^{d} \left( p_0, \vec{p} \right) = \frac{1}{2 \omega_{R} (\vec{p}) \left[ \omega_{R} (\vec{p}) - p_0 - i \epsilon \right] - \Sigma_{R} (p_0, \vec{p}) } \; , \label{dressed-roper-prop}
\end{align}
where $\omega_{R} (\vec{p}) = \big[ | \vec{p} |^{2} + m_{R}^{2} \big]^{1/2} $, $m_R$ and $\Sigma_{R} (p_0, \vec{p})$ is the self-energy of the Roper in the particle-dimer framework. This self-energy should again be considered in the finite volume, i.e. according to Eq.~\eqref{FV-formalism}. The poles of the finite-volume propagator obey 
\begin{align}
2 \omega_{R} (\vec{p}) \left[ \omega_{R} (\vec{p}) - p_0  \right] - \Sigma_{R}^L (p_0, \vec{p}) = 0 \quad \Rightarrow \quad m_R - E - \frac{1}{2 m_{R}} \Sigma_{R}^L (E) = 0 \; ,  \label{Energy-levels-FV-Dimer} 
\end{align}
where the rest frame condition ($p_0 = E$, $\vec{p} = 0$) has been used to obtain a relation for the finite-volume energy levels. This equation on the right-hand side in~\eqref{Energy-levels-FV-Dimer} must be solved next. For an easier overview, only the first two diagrams in Fig.~\ref{fig:diags2} are considered, i.e. the diagrams with intermediate $N \pi$ and intermediate $N \sigma$. Starting with the diagram with internal nucleon and pion, one obtains the self-energy contribution 
\begin{align}
\Sigma_{N \pi}^L \left( E \right) = \frac{f_2^2}{4 \pi^{3/2} E L} \mathcal{Z}_{00} \left( 1, \tilde{q}^2 (E) \right)  \; , \quad \text{with} \quad \tilde{q}^2 (E) = \left( \frac{L}{2 \pi} \right)^2  \frac{\lambda \left( E^2, m_{N}^2, M_{\pi}^2  \right)}{4 E^2 } \; ,  \label{J-pi-N-FV-result}
\end{align}
where $\mathcal{Z}_{00}$ is the L{\"uscher} function~\cite{Luscher:1990ux}. The more interesting case is the diagram with internal nucleon and $\sigma$-dimer, since the $\sigma$ is not an asymptotic final state, but can decay into two pions. To include three-particle dynamics, the $\sigma$-dimer propagator needs to be dressed, i.e. 
\begin{align}
D_{\sigma} \left( p \right) = - \frac{1}{\alpha_{\sigma}^{} M_{\sigma}^2 + \Sigma_{\sigma}^{} (p) } \; , \quad \text{with} \quad \Sigma_{\sigma}^{} \left( p \right)  =  2 h_2^2 \int \frac{d^4 k}{(2 \pi)^4 i} S_{\pi} (p - k) S_{\pi} (k) \; . 
\end{align}
The self-energy contribution to the Roper resonance in the finite volume is then given by the double sum 
\begin{align}
\begin{split}
\Sigma_{N \sigma}^{L} \left( E \right) &= - f_4^2 \frac{1}{L^3} \sum_{\vec{k}} \frac{1}{2 \omega_{N} (\vec{k})} \left \lbrace \alpha_{\sigma}^{} M_{\sigma}^2 \phantom{ \sum_{\vec{l}} } \right. \\
&\phantom{=} \; \left.  + 2 h_2^2  \frac{1}{L^3} \sum_{\vec{l}} \, \frac{1}{ 4 \omega_{\pi} (\vec{k} - \vec{l}) \omega_{\pi} (\vec{l})   \big[ \omega_{\pi}(\vec{k} - \vec{l}) + \omega_{\pi} (\vec{l}) + \omega_{N} (\vec{k}) - E  \big]}  \right \rbrace^{-1} . \label{Nsigma-double-sum}
\end{split}
\end{align} 
It can be observed that $\Sigma_{N \sigma}^{L} \left( E \right)$ depends on two coupling constants: the coupling $f_4$, which describes the interaction between the Roper, the nucleon and the $\sigma$-dimer, and the coupling $h_2$, which describes the interaction between the $\sigma$-dimer and two pions. Both constants together determine the positions of the interacting energy levels. 

\begin{figure}[t]
	\includegraphics[width=0.49\linewidth]{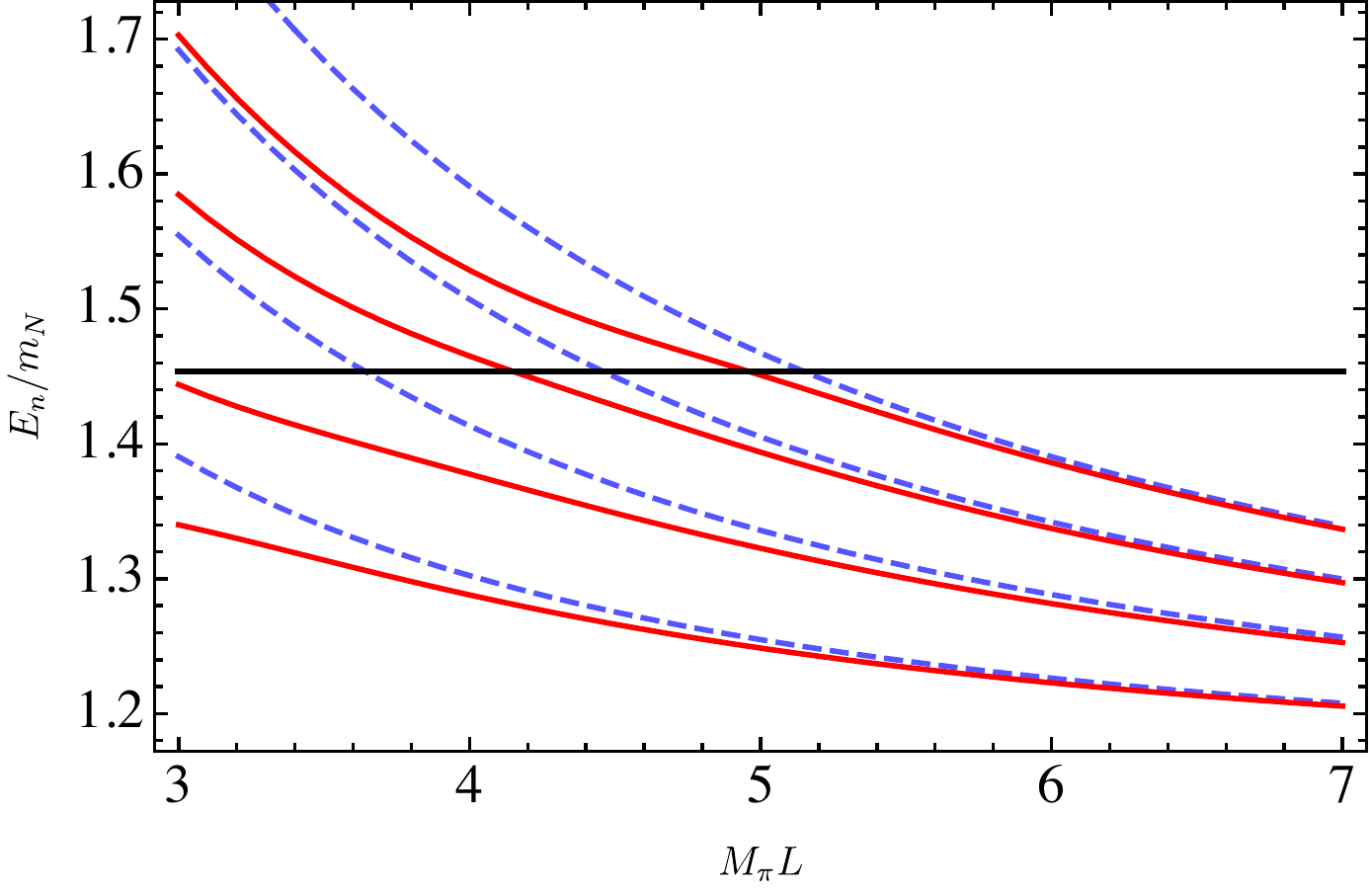}
	~
	\includegraphics[width=0.49\linewidth]{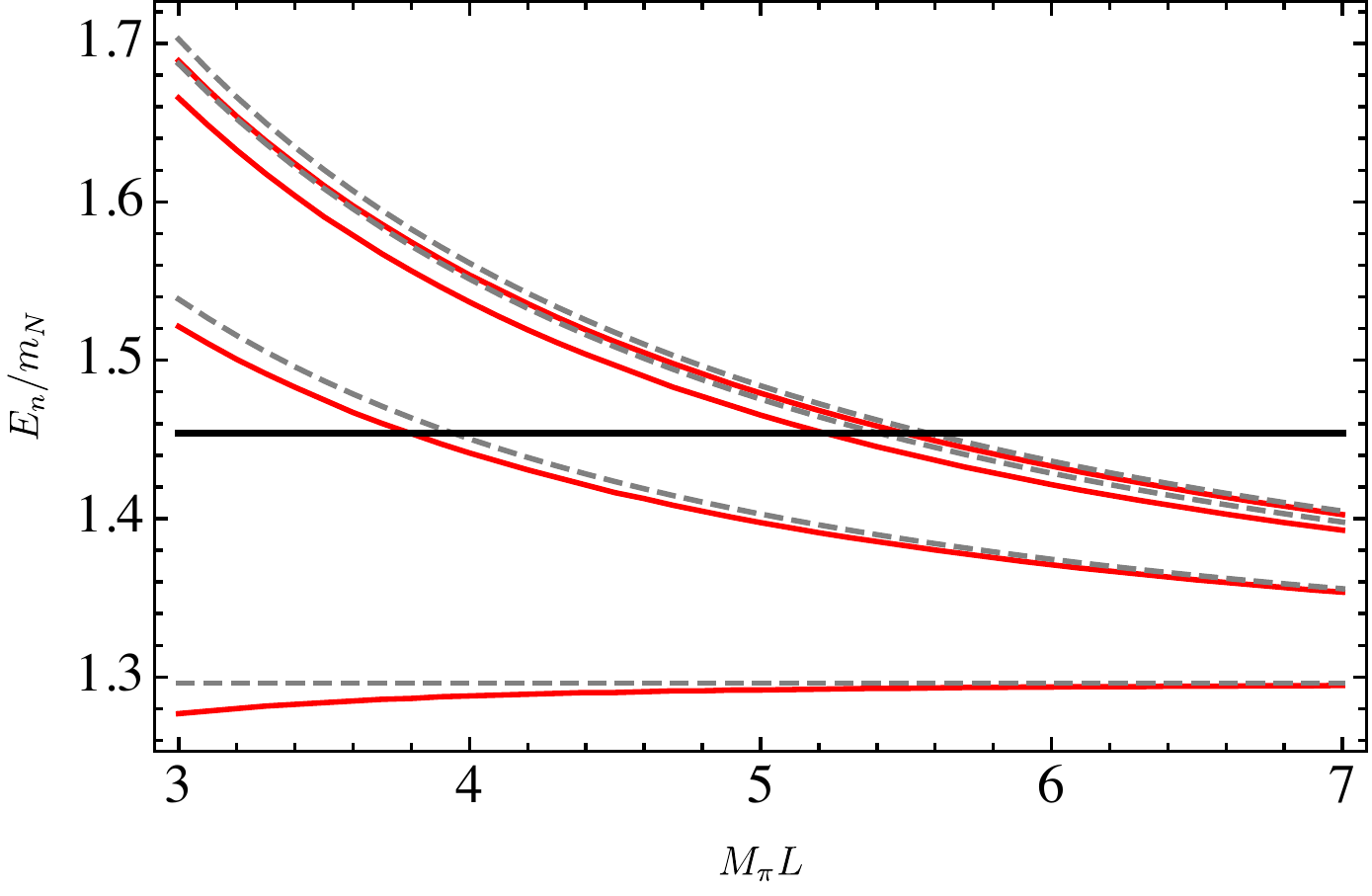}
	\caption{
		\textbf{Left}: Energy levels of the Roper resonance in the particle-dimer picture for different box sizes $L$ considering only pion and nucleon as intermediate states. Red solid lines display the numerical results for the interacting levels and blue dashed lines the free (non-interacting) energy levels of the pion and nucleon. The thick black line marks the mass of the Roper resonance, i.e. $m_R / m_N$.
		\textbf{Right}: Energy levels of the Roper resonance in the particle-dimer picture for different box sizes $L$ considering only nucleon and $\sigma$-dimer as intermediate states. Red solid lines display the numerical results for the interacting levels and gray dashed lines the free (non-interacting) lowest-lying three-particle $N \pi \pi$ energy levels. The thick black line marks the mass of the Roper resonance, i.e. $m_R / m_N$.}
	\label{fig:En-levels-dimerformalism}
\end{figure}

Now, the different self-energy contributions can be calculated numerically. Analogously to the BChPT analysis, let us first consider only the contribution from internal nucleons and pions, $\Sigma_{N \pi}^L \left( E \right)$. One arrives at the left picture of Fig.~\ref{fig:En-levels-dimerformalism}, which shows a similar behavior of the energy levels as in the BChPT case. In fact, both energy spectra agree very well with each other. Avoided level crossing is again seen around the critical value (thick black line in Fig.~\ref{fig:En-levels-dimerformalism}) and the interacting energy levels approach the free levels for larger box sizes $L$. This is a remarkable result, considering that the non-relativistic EFT framework is much simpler than the Lagrangian from BChPT~\cite{Severt:2020jzc} and does not include any spin structure or full Lorentz-symmetry. The agreement of the energy levels between the different approaches is a profound argument to continue the non-relativistic EFT investigation. Next, let us consider the contribution $\Sigma_{N \sigma}^{L} \left( E \right)$ from the nucleon and the $\sigma$-dimer field. The result of the numerical calculation is given in the right panel of Fig.~\ref{fig:En-levels-dimerformalism}. This spectrum does not show two-particle final states, but three-particle final states. The interacting energy levels lie very close to the non-interacting $N \pi \pi$ levels and approach them asymptotically for large box sizes. Interestingly, it is observed that avoided level crossing as a signature of a presence of a resonance is widely suppressed. Thus, the spectrum needs to be investigated further in the future to find out how the couplings $f_4$ and $h_2$ influence the behaviour of the energy levels.

\section{Summary and outlook}

In our previous work Ref.~\cite{Severt:2020jzc}, we demonstrated how the Roper resonance can be treated within BChPT considering two-particle final states only. In the next step we introduced a non-relativistic particle-dimer formalism for the Roper, which can reproduce the $N \pi$-spectrum from BChPT accurately, but also allows one to incorporate proper three particle $N \pi \pi$ dynamics.  

Some more work on the particle-dimer picture for the Roper resonance is needed: The obtained spectrum of the Roper for intermediate nucleon and $\sigma$-dimer states should be analyzed further for different couplings and, analogously to the calculations in~\cite{Severt:2020jzc}, both $N \pi$ and $N \sigma$ interactions should be turned on to obtain the energy levels of the coupled-channel system. Also, the spectrum of the Roper for internal pions and $\Delta$-dimer fields should be calculated. Additionally, the application of the particle-dimer framework to different hadronic systems, for example the $a_{1} (1260)$ resonance (see e.g.~\cite{Mai:2021nul}), might be considered in the future.

\section*{Acknowledgments}

The speaker wants to thank the organizers of the conference for the opportunity to give this talk. Additionally, the speaker expresses his gratitude to his collaborators \textit{Ulf-G. Mei{\ss}ner}, \textit{Maxim Mai} and \textit{Akaki Rusetsky} for their significant contributions to this project. This work was supported by the Deutsche Forschungsgemeinschaft (DFG, German Research Foundation, Project number~$196253076$ - TRR~$110$).


\end{document}